\documentclass[conference]{IEEEtran}
\IEEEoverridecommandlockouts
\usepackage{cite}
\usepackage{amsmath,amssymb,amsfonts}
\usepackage{algorithmic}
\usepackage{graphicx}
\usepackage{textcomp}
\usepackage{xcolor}
\usepackage{enumitem}
\usepackage{dsfont}
\usepackage{hyperref}
\usepackage{amsmath}
\newcommand{\norm}[1]{\left\lVert#1\right\rVert}
\def\BibTeX{{\rm B\kern-.05em{\sc i\kern-.025em b}\kern-.08em
    T\kern-.1667em\lower.7ex\hbox{E}\kern-.125emX}}
\begin{document}

\title{In-Context Learning for MIMO Equalization Using Transformer-Based Sequence  Models\\

\thanks{
	Matteo Zecchin and Osvaldo Simeone are with the King’s Communications, Learning \& Information
	 Processing (KCLIP) lab within the Centre for Intelligent Information Processing Systems (CIIPS), Department of Engineering, King’s College London, London WC2R 2LS, U.K. (e-mail: matteo.1.zecchin@kcl.ac.uk;
	 osvaldo.simeone@kcl.ac.uk)
	Kai Yu is with the School of Electronic Science and Engineering, Nanjing
	 University, Nanjing, China, 210023.(e-mail: kaiyu@smail.nju.edu.cn).
	
	The work of M. Zecchin and O. Simeone was supported by the European Union’s Horizon Europe project CENTRIC (101096379). The work of O. Simeone was also supported by the Open Fellowships of the EPSRC (EP/W024101/1) by the EPSRC project  (EP/X011852/1), and by Project REASON, a UK Government funded project under the Future Open Networks Research Challenge (FONRC) sponsored by the Department of Science Innovation and Technology (DSIT).
	
	O. Simeone produced the original idea and supervised the work and the writing; K. Yu prepared some part of the  code used for preliminary results, and contributed  to the writing; M. Zecchin completed and extended the code, producing all the experimental results, and had the main role in drafting the paper.}
}
\author{
    \IEEEauthorblockN{Matteo Zecchin, Kai Yu and Osvaldo Simeone}
}

\maketitle

\begin{abstract}
Large pre-trained sequence models, such as transformer-based architectures, have been recently shown to have the capacity to carry out in-context learning (ICL).  In ICL,  a decision on a new input is made via a direct mapping of the input and of a few examples from the given task, serving as the task's context, to the output variable. No explicit updates of the model parameters are needed to tailor the decision to a new task. Pre-training, which amounts to a form of meta-learning, is based on the observation of examples from several related tasks. Prior work has shown ICL capabilities for linear regression. In this study, we leverage ICL to address the inverse problem of multiple-input and multiple-output (MIMO) equalization based on a context given by pilot symbols. A task is defined by the unknown fading channel and by the signal-to-noise ratio (SNR) level, which may be known. To highlight the practical potential of the approach, we allow the presence of quantization of the received signals. We demonstrate via numerical results that transformer-based ICL has a threshold behavior, whereby, as the number of pre-training tasks grows, the performance switches from that of a minimum mean squared error (MMSE) equalizer with a prior determined by the pre-trained tasks to that of an MMSE equalizer with the true data-generating prior.  

\end{abstract}


\section{Introduction}

\label{sec:intro}
\begin{figure}[!t]
	\centering
	\includegraphics[width=0.44\textwidth]{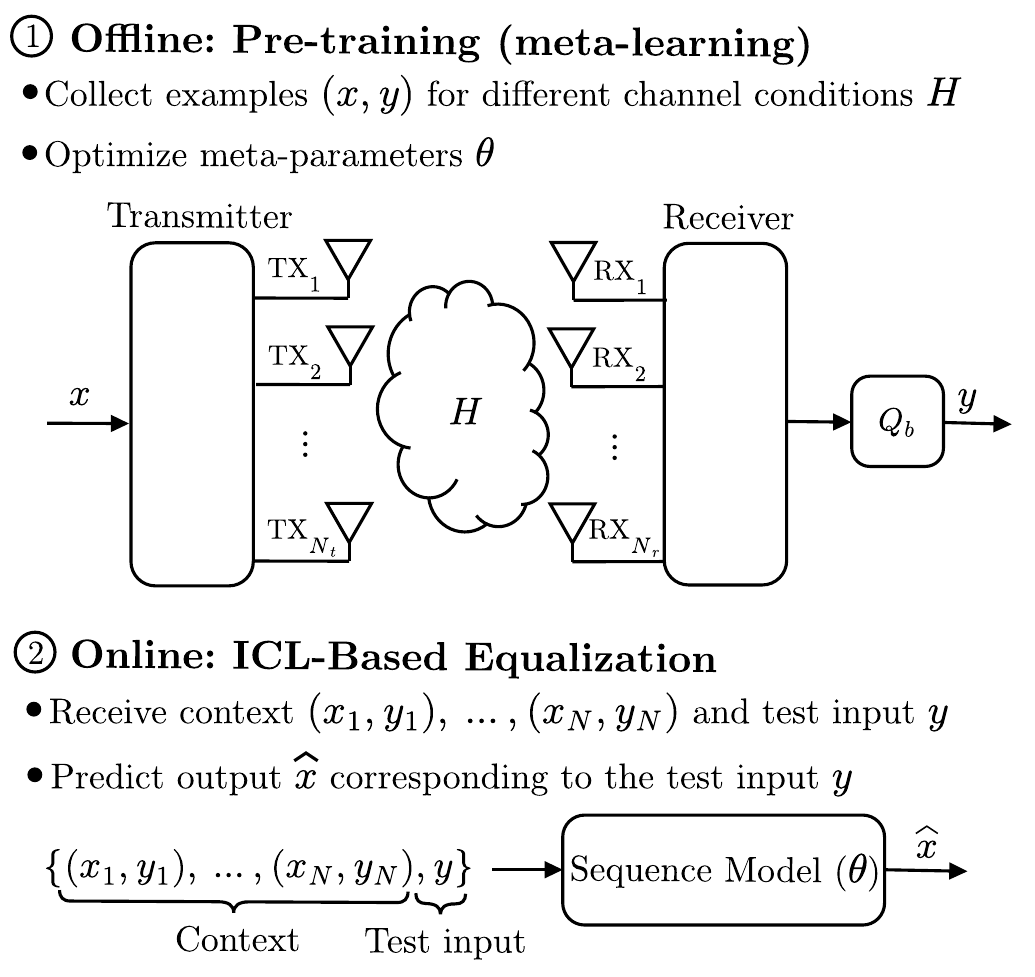}
	\centering
    \vspace{-0.1cm}
	\caption{Illustration of in-context learning for MIMO equalization. }
  \vspace{-0.4cm}
	\label{fig1}
\end{figure}
\noindent \emph{Context and motivation:} With the advent of disaggregated radio access networks (RAN), artificial intelligence (AI) models are increasingly expected to play a central role in next-generation wireless systems \cite{groen2023implementing}. For AI to be successfully deployed on the RAN, an important requirement is the ability to quickly adapt to changing environmental conditions based on limited contextual information \cite{simeone2020learning,chen2023learning}, possibly with the help of powerful simulation intelligence tools such as digital twins \cite{dong2022survey,ruah2023bayesian}. As a notable example, an AI-based wireless receiver should be able to update its internal operation on the basis of limited pilots, so as to ensure satisfactory performance despite time-varying channel conditions \cite{raviv2023modular}. AI-based receivers are particularly well suited for situations characterized by algorithmic deficits, i.e., for communication scenarios in which optimal algorithms are unknown or hard to implement, as is the case with non-linear impairments in the transceiver chain \cite{simeone2018very}.

\noindent\emph{Meta-learning}, or learning to learn, offers a general framework for the design of AI models that can efficiently adapt on the basis of a few examples \cite{chen2023learning,simeone2022machine}. However, conventional meta-learning schemes require the implementation of explicit optimization procedures for the update of the model parameters, causing potential issues with computational complexity and lack of robustness to the selection of hyperparameters. Recently, \emph{large pre-trained sequence models}, also known as large language models (LLMs), have emerged as an alternative, due to their capacity to implement  \emph{in-context learning (ICL)} \cite{dong2022survey,min2022metaicl,akyurek2022learning,zhang2023trained}.  This paper aims at exploring, and understanding, the potential of such models for the specific problem of equalization over non-linear multiple-input multiple-output (MIMO) channels. Concurrent work \cite{rajagopalan2023transformers} has proposed the same basic concept, and we will discuss below some differences between the two studies.

\noindent \emph{In-context learning:}  ICL can be viewed as a form of meta-learning requiring no explicit model updates, a process also known as \emph{mesa-learning} \cite{Oswald2023UncoveringMA}. Prompted with a description of the task in the form of example input-output pairs, referred to as \emph{context}, a sequence model exhibits ICL capabilities if it can directly assign an output for a given input without requiring fine-tuning. Recent papers \cite{zhang2023trained, akyurek2022learning} demonstrated theoretically and experimentally the capacity of large sequence models to implement ICL for special classes of functions, with a specific focus on linear models. ICL capabilities are acquired by \emph{pre-training} the models on data for a number of tasks that are expected to be related to the tasks to be encountered at run time.

In particular, reference \cite{raventos2023pretraining} provided empirical evidence of a threshold behavior for linear regression models. When the number of pre-training tasks is below a threshold, an ICL-based model exhibits similar performance to a system equipped with knowledge of the true linear regressors for the tasks seen during pre-training. In contrast, for sufficiently many pre-training tasks, the performance tends to that of an ideal Bayesian predictor that knows the true underlying distribution of the linear regressors.

\noindent \emph{Contributions:}  In this work, we propose the use of pre-trained sequence models for the implementation of equalization over non-linear MIMO channels via ICL. As seen in Fig. 1, given a context in the form of a number of pilot symbols and a new received signal, a pre-trained sequence model is leveraged to directly produce an estimate of the transmitted signal by running a forward pass of the model. Our main contributions are as follows.
\begin{itemize}[noitemsep, topsep=0pt,wide=0pt]
	\item 
	We present a framework for ICL-based equalization in the presence of non-linear MIMO channels. Unlike prior works \cite{zhang2023trained, akyurek2022learning} on ICL, the setting under study here address an \emph{inverse} problem, rather than a more conventional regression problem.
	\item
	Inspired by \cite{raventos2023pretraining}, we show numerically that, even in the given inverse problem setting, ICL presents a threshold behavior as a function of the number of pre-training tasks, switching between a discrete minimum mean squared error (MMSE) equalizer using as prior the channels observed during pre-training to an optimal MMSE equalizer based on the true channel distribution.
	\item We compare the performance of ICL to conventional meta-learning for equalization \cite{park2020learning}, illustrating the superior performance of ICL in the case of short pilot sequences.
\end{itemize}
In this regard, as compared to the recent concurrent work \cite{rajagopalan2023transformers}, (\emph{i}) we explore \emph{non-linear} MIMO channels; (\emph{ii}) we reveal the mentioned threshold behavior as a function of the number of pre-training tasks; and  (\emph{iii}) we provide a performance comparison with model-agnostic meta-learning (MAML) \cite{finn2017model}.

\section{System Model}

\subsection{Signal Model}

\label{sec:ch_eq}
We consider the problem of channel equalization for $N_t\times N_r$ MIMO systems affected by additive white complex Gaussian noise channels and subject to non-linear impairments in the receiving chain. A \emph{channel equalization task} is characterized by a tuple  $\tau=(H,\sigma^2)$, which consists of a $N_r\times N_t$ complex-valued channel matrix $H$  and of the variance $\sigma^2$ of the channel noise vector.  The channel input, given by vector $x\in\mathbb{C}^{N_t}$, is assumed to be drawn uniformly at random from some known constellation set $\mathcal{X}$, and we normalize the overall average transmit power to 1, i.e., we set $\mathbb{E}[\|x\|^2]=1$. Accordingly,  the per-receive antenna average signal-to-noise ratio (SNR) is given by the inverse of the channel variance \begin{equation}\label{snr}
	\text{SNR} = \frac{\mathbb{E}[\|x\|^2]}{\sigma^2}=\frac{1}{\sigma^{2}}.
\end{equation} 

Furthermore, the received signal is given by \begin{equation}
	y = Q_b(Hx + z),
	\label{eq:received_quantized}
\end{equation}
where  $Q_b(\cdot)$ denotes a quantizer with resolution $b$ bits that is applied separately to in-phase and quadrature components. 

\subsection{ICL-Based Equalization}
For an equalization task $\tau$, we write as $P_{y,x|\tau}$ the joint distribution of the uniformly distributed inputs $x$ and of the received signal $y$ in (\ref{eq:received_quantized}). Tasks are assumed to be characterized by an unknown distribution \begin{equation} \label{eq:indep}P_\tau=P_{H,\sigma^2}=P_{H}P_{\sigma^2} \end{equation} over the channel fading matrix $H$ and channel noise level $\sigma^2$. Note that the independence of channel and SNR level implied by (\ref{eq:indep}) is not necessary for what follows, but it appears to be a practically reasonable assumption. The goal of equalization is to estimate the input $x$ for a given received signal $y$. 

To this end, the ICL-based equalizer, which is unaware of the identity of the current task $\tau$, only has access to a \emph{context} given by a sequence of $N$ independent and identically distributed (i.i.d.) pilots \begin{equation}\mathcal{D}_\tau=\{(x_i,y_i)\}^N_{i=1}\sim P_{y,x|\tau}^{\otimes N}.\end{equation} With this information, the equalizer wishes to assign to a new received signal $y$ from the same task $\tau$, the corresponding input $x$, which is unknown. The test pair $(y,x)\sim P_{y,x|\tau}$ is independent of the examples in context set $\mathcal{D}_\tau$. 

Overall, the \emph{ICL-based equalizer} produces an estimate $\hat{x}$ based on the parameterized mapping 
\begin{align}
	\hat{x}=\hat{x}_\theta(\mathcal{D}_\tau,y),
	\label{eq:ICLeq}
\end{align} where $\theta$ is a vector of parameters defining the equalizer. We emphasize that  the equalizer (\ref{eq:ICLeq}) is not aware of the task $\tau$, i.e., it does not know the current channel realization $H$ and the SNR level (\ref{snr}). The performance of the equalizer is quantified by the mean squared error (MSE)
\begin{align}
	\textrm{MSE}_{\tau}(\theta)=\mathbb{E}_{(y,x)\sim P_{y,x|\tau}}\left[\norm{\hat{x}_{\theta}(\mathcal{D}_\tau,y)-x}^2\right]
	\label{eq:mmse_problem}
\end{align}where $\mathbb{E}_{(y,x)\sim P_{y,x|\tau}}[\cdot]$ represents the expectation with respect to the task-specific distribution $P_{y,x|\tau}$.

\subsection{Pre-Training (Meta-Learning)}
\label{sec:icl_eq}
As shown in Fig. \ref{fig1}, during an initial offline phase, the parameters $\theta$ of the ICL-based equalizer in (\ref{eq:ICLeq}) are optimized based on the observation of context and test data from multiple tasks $\tau$. The $M$ pre-training tasks are generated i.i.d. from distribution $P_\tau$, and we denote as \begin{align}\label{eq:tasks} \{\tau_m=(H_m,\sigma^2_m)\}_{m=1}^M \sim P_\tau^{\otimes M} \end{align}
the corresponding task parameters.  

For each pre-training task $\tau_m$, during pre-training, we have access to context $D_{\tau_m}$ and test pair $(y_m,x_m)$. Note that, for pre-training, unlike in the online phase, the equalization target $x_m$ is known. The pre-training goal is to minimize the MSE (\ref{eq:mmse_problem}), averaged over the given pre-training tasks $M$, with respect to the parameters of the ICL-based equalizer $\theta$. Accordingly, the \emph{training loss} function is given by the sum 
\begin{align}\label{loss_pretrain}
	L(\theta)&=\sum_{m=1}^M \textrm{MSE}_{\tau_m}(\theta)\nonumber \\
 &= \sum_{m=1}^M \mathbb{E}_{(y,x)\sim P_{y,x|\tau_m}}\left[\norm{\hat{x}_{\theta}(\mathcal{D}_{\tau_m},y_m)-x_m}^2\right].
\end{align}

Comparing the problem formulation (\ref{loss_pretrain}) with recent papers including \cite{akyurek2022learning,zhang2023trained,Garg2022WhatCT}, we observe the following important difference. The underlying data-generation model in most existing works assumes a functional dependence $x=f(y)$ between input and output, respectively $y$ and $x$ in our notation, often considered to be a linear function. In contrast, here we study an inverse problem, in which data are generated according to the forward model (\ref{eq:received_quantized}), and the goal is to invert this relationship to estimate $x$ from $y$.  As mentioned in Sec. \ref{sec:intro}, the same problem was studied in the concurrent work \cite{rajagopalan2023transformers} without accounting for the presence of quantization and without pointing to the threshold behavior related to the baselines introduced in the next section.

\section{Baselines}
\label{sec:baselines}
In this section, we introduce relevant baselines for the channel equalization problem. Unlike the concurrent work \cite{rajagopalan2023transformers}, we follow the insights of paper \cite{raventos2023pretraining}, which shows, for linear regression, that ICL can exhibit a threshold behavior as a function of the number of pre-training tasks. Our experiments in the next section will reveal a similar behavior for the equalization problem.

\subsection{MMSE Equalizer for a Known Task }\label{subsec:ideal}
In an ideal situation, the equalizer knows the current task $\tau$, consisting of channel matrix $H$ and channel noise power $\sigma^2$, and thus also the joint distribution $P_{y,x|\tau}$ of received signal $y$ and input $x$. For this scenario, the minimizer of the MSE in (\ref{eq:mmse_problem}) is given by the minimum MSE (MMSE) estimator
\begin{align}
	x_\tau^*=\mathrm{E}_{x\sim P_{x|y,\tau}}[x]= \sum_{x\in\mathcal{X}} x P_{x|y,\tau},
	\label{eq:mmse}
\end{align}
where $P_{x|y,\tau}$ is the posterior distribution $P_{x|y,\tau}=P_{y,x|\tau}/P_{y|\tau}$, where $P_{y|\tau}=\sum_{x\in\mathcal{X}} P_{y,x|\tau}$ is the marginal distribution of the received signal. 

The estimator (\ref{eq:mmse}) requires non-linear operations on the received signal $y$. Therefore, in virtue of its simplicity, we will also consider a linear MMSE (LMMSE) estimator obtained under simplifying assumptions. Specifically, if one assumes that the input $x$ is Gaussian with distribution $x\sim  \mathcal{CN}(0,1/N_r I_{N_r})$ and if we neglect the presence of the quantizer, the MMSE solution for a task $\tau$ can be computed  as 
\begin{align}
	\hat{x}^{\text{*,lin}}_\tau=(2\sigma^2 I+H^\text{H}H)^{-1}H^\text{H}y.
	\label{eq:linearMMSE}
\end{align} Note that, due to the misspecification of the model underlying the equalizer (\ref{eq:linearMMSE}), the LMMSE solution (\ref{eq:linearMMSE}) is suboptimal when the task $\tau$ is known.

\subsection{MMSE Equalizer with Known SNR and a Given Channel Distribution}\label{subsec:knowndist}

Suppose now that the estimator is given the SNR level, but it only has access to a prior distribution $Q_H$ for the channel matrix $H$,  as well as to  the context  $\mathcal{D}_\tau$ for the new task $\tau$. As detailed below, we will consider two different priors $Q_H$ corresponding to situations in which the equalizer knows the true data-generating distribution or else it only knows the channels generating the data used in the pre-training phase.

With the given information, the posterior distribution of the channel matrix $H$ for the test task $\tau$ is given as 
\begin{align}
	Q_{H|\mathcal{D}_\tau}\propto Q_H \prod^N_{i=1}  P_{y_i|x_i,\tau},
	\label{posterior}
\end{align}with the likelihood $P_{y_i|x_i,\tau}$ describing the model (\ref{eq:received_quantized}). 
Based on the posterior distribution (\ref{posterior}), the MMSE solution is given as the average 
\begin{align}
	\hat{x}^{\mathrm{dist}}_\tau=\mathrm{E}_{H\sim Q_{H|\mathcal{D}_\tau}} [ \mathrm{E}_{x\sim P_{x|y,\tau}}[x]].
	\label{eq:BayesMMSE}
\end{align}

As anticipated, we consider two different choices for the channel prior $Q_H$.  \begin{itemize}
    \item \emph{Known channel distribution}: In this first setting, the equalizer is  given the true channel distribution $P_H$ (see (\ref{eq:indep})), which is described by the prior choice  $Q_H=P_H$. 
    \item \emph{Known pre-training channels}: In the second scenario, the equalizer is only aware of the channel realizations $\mathcal{H}=\{H_m\}_{m=1}^M \sim P_H^{\otimes M}$ in (\ref{eq:tasks}), which determine the received signals during pre-training. Accordingly, the  prior $Q_H$ is selected as a discrete uniform distribution over set $\mathcal{H}$, which is denoted as $Q_H=U_\mathcal{H}$.
\end{itemize} 

The MSE obtained under the first scenario serves as an optimistic benchmark for the ICL-based equalizer, which does not have access to the true distribution of the channel $H$. In contrast, the second setting describes a possible solution strategy for the ICL-based equalizer. According to this strategy, the equalizer estimates the channels $\{H_m\}_{m=1}^M$ based on the context data during pre-training; and then uses these channels to determine a prior $Q_H$ to be used for the MMSE estimate (\ref{eq:BayesMMSE}). While this benchmark assumes known pre-training channels -- information that is not given to the ICL-based equalizer --, its MSE performance does not set a limit to the MSE achievable by the ICL-based equalizer. In fact, given the available pre-training data, the ICL-based equalizer can effectively attempt to extrapolate mappings (\ref{eq:ICLeq}) that mimic more closely the first setting corresponding to the known channel-distribution performance, i.e., $Q_H=P_H$.

\section{Transformer-based Sequence Model for Equalization}

\label{sec:tf_equalizer}

\begin{figure}[!t]
	\centering
	\includegraphics[width=0.35\textwidth]{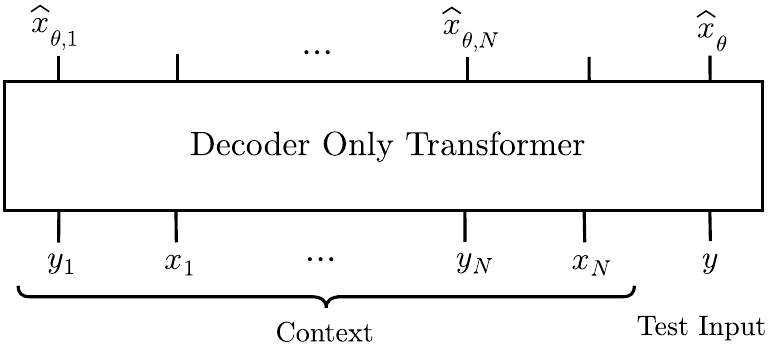}
	\centering
    \vspace{-0.15cm}
	\caption{Decoder-only transformer for ICL-based MIMO equalization.}
	\label{fig2}
  \vspace{-0.3cm}
\end{figure}
As discussed in Sec. \ref{sec:icl_eq}, an ICL-based equalizer implements function (\ref{eq:ICLeq}) that maps the context $\mathcal{D}_\tau$ and received signal $y$ to the estimated channel input $\hat{x}_{\theta}(\mathcal{D}_\tau,y)$. As in \cite{Garg2022WhatCT}, this mapping is realized by a \emph{decoder-only autoregressive transformer} architecture \cite{Garg2022WhatCT}.  


As illustrated in Fig. \ref{fig2}, the transformer-based equalizer takes as input the sequence 
\begin{equation}
(\mathcal{D}_{\tau},y)=(y_1,x_1,y_2,x_2,...,y_N,x_N,y).
\label{eq:input_seq}
\end{equation}
corresponding to the concatenation of the context $\mathcal{D}_{\tau}$ and the test point $y$. To this input, it applies the following steps in order to produce the estimate (\ref{eq:ICLeq}). These steps are parameterized by several trainable parameters, which constitute the parameter vector $\theta$.

\noindent \emph{Linear embedding:}  The first step consists in an embedding operation $E(\cdot)$ that maps every element of the input sequence (\ref{eq:input_seq}) into a vector of common dimension $D_e$. To this end, all the vectors of transmitted symbols $x\in\mathbb{C}^{N_t}$ are mapped into corresponding real-valued vectors $\tilde{x}\in\mathbb{R}^{2N_t}$ by concatenating  real and imaginary components as 
\begin{align}
    \tilde{x}=[\Re(x),\Im(x)]\in\mathbb{R}^{2N_t},
\end{align}
and the same transformation is applied to the received vectors $y$ to obtain the real-valued vectors  $\tilde{y}\in\mathbb{R}^{2N_t}$. Prior to embedding, we zero-pad the shorter of the two sets of vectors $\tilde{x}$ and $\tilde{y}$ so that they all have the same dimension $D_s=2\max\{N_t,N_r\}$. Accordingly, in the following, the notations $\tilde{x}$ and $\tilde{y}$ refer to $D_s\times 1$ real-valued vectors.

The post-embedding sequence $E=E(\mathcal{D},y)\in \mathbb{R}^{D_e\times (2N+1)}$ is obtained by multiplying each vector -- be it a transmitted vector $\tilde{x}$ or a received vector $\tilde{y}$ -- with a trainable embedding matrix $M_e\in \mathbb{R}^{D_e\times D_s}$, i.e., \begin{equation}
E=E(\mathcal{D}_{\tau},y)=(M_e\tilde{y}_1,M_e\tilde{x}_1,...,M_e\tilde{y}_N,M_e\tilde{x}_N,M_e\tilde{y}).
\label{eq:emb_input_seq}
\end{equation} 
 Each column of matrix $E$ corresponds to a \emph{token}.

\noindent \emph{Multi-head self-attention:} The embedded sequence $E$ is then processed by the repeated application of a multi-head attention mechanism across $L$ layers \cite{Vaswani2017AttentionIA}. Each layer $l$ produces a sequence of $2N+1$ transformed tokens $E^l$ for $l=1,...,L$. Intuitively, as shown in Fig. \ref{fig2}, the tokens corresponding to each received signal vector $y$ -- and the corresponding embedding version $\tilde{y}$ -- should provide enough information to enable an effective estimate of the corresponding transmitted signal $x$. 

Each $l$-th layer takes as input the sequence of tokens $E^{l-1}$ from the previous layer with $E^0=E$. It applies $H$ attention ``heads'', which are combined in order to produce the input $E^l$ for the next layer. As we will detail, for the last, $L$-th, layer, the output $E^L$ is used to carry out equalization.  { The attention operation of each layer $l$ is defined by $3H$ trainable weight matrices, namely the key matrices $W^K_h\in\mathbb{R}^{D_w\times D_e}$, the query matrices $W^Q_h \in \mathbb{R}^{D_w\times D_e}$, and the value matrices $W^V_h \in \mathbb{R}^{D_v\times D_e}$ for all heads $h=1,...,H$. As in \cite{Vaswani2017AttentionIA}, we set $D_w=D_v=D_e/H$.}

For each head $h$ at layer $l$, softmax self-attention produces the  $2N+1$  modified tokens
\begin{equation}\label{eq:attention}
\begin{aligned}
B_h^l&= W^V_hE^{l-1} \cdot \text{softmax}\left(\frac{(W^K_h E^{l-1})^T(W^Q_hE^{l-1})}{\sqrt{D_k}}\right),
\end{aligned}
\end{equation}where we have used a notation similar to \cite{zhang2023trained}, whereby the softmax function is applied  column-wise.  By \eqref{eq:attention}, each output token is a convex combination of the input tokens in sequence $E^{l-1}$ with weights dictated by the softmax function with logits given by the inner products between corresponding queries, i.e., columns of matrix $W^Q_hE^{l-1}$, and keys, i.e., columns of matrix  $W^K_h E^{l-1}$ \cite{turner2023introduction}. 

The output of all $h$ heads are concatenated token by token, and linearly projected, also token by token, into a sequence of $2N+1$ tokens of dimension $D_e \times 1 $ as  
\begin{equation}
A^l=(W^O)^T\left[ B_1^l,B_2^l,\cdots,B_H^l\right],
\end{equation}
where $W^O \in \mathbb{R}^{HD_v \times D_{e}}$ is another trainable weight matrix. Finally, each token is passed in parallel through a two-layer feed-forward neural network with residual connections to produce the output 
\begin{equation}
E^{l}=W_1\delta(W_2\beta(A^l+E^{l-1}))+A^l+E^{l-1},
\end{equation}
where $W_1\in \mathbb{R}^{D_e \times D_{f}}$ and $W_2\in \mathbb{R}^{D_{f}\times D_e}$ are trainable weight matrices with $D_{f}$ denoting the number of hidden neurons; $\beta(\cdot)$ is the layer norm operation \cite{Ba2016LayerN}; and $\delta(\cdot)$ is the Gaussian error linear unit (GeLU) activation function \cite{Hendrycks2016GaussianEL}.

\noindent \emph{Equalization:}  Finally, the output tokens $E^L$ produced by the last layer are passed through a trainable linear softmax classifier with a number of outputs equal to the constellation elements to carry out equalization. As illustrated in Fig. \ref{fig2} the output corresponding to the last token is taken as the final output (\ref{eq:ICLeq}) of the equalizer. 

\section{Results}
In this section we investigate the performance of the ICL-based equalizer introduced in Sec. \ref{sec:tf_equalizer} in comparison with the relevant baselines reviewed in Sec. \ref{sec:baselines}.  The code is available at the link \url{https://github.com/kclip/ICL-Equalization}.

\emph{Set-up:} We consider a $2\times 2$ MIMO system, i.e., $N_t=N_r=2$, in which the channel input $x$ is sampled from a 4-QAM constellation $\mathcal{X}$, and the received signal is quantized using a mid-rise $b$ bit uniform quantizer with range $[-4,4]$. We assume that the channel distribution $P_H$ prescribes i.i.d. complex Gaussian variables $H_{i,j}\sim\mathcal{CN}(0,1)$, and that the noise variance $\sigma^2$ is uniformly distributed within the interval $[\sigma^2_{\min} [\mathrm{dB}],\sigma^2_{\max} [\mathrm{dB}]]$ for some boundary values $\sigma^2_{\min}$ and $\sigma^2_{\max}$ to be specified.

The ICL-based equalizer is instantiated using a transformer architecture consisting of $L=2$ attention layers with $H=4$ heads each and with an embedding dimension $D_e=64$. Unless explicitly mentioned otherwise, we set $b=4$, and the model is trained on a pre-training task set (\ref{eq:tasks}) consisting of $M=4096$ tasks. For each task $\tau$, a context $\mathcal{D}_\tau$ contains $N=20$ labelled examples.

\begin{figure}
	\centering
    \vspace{-0.2cm}
	\includegraphics[width=0.4\textwidth]{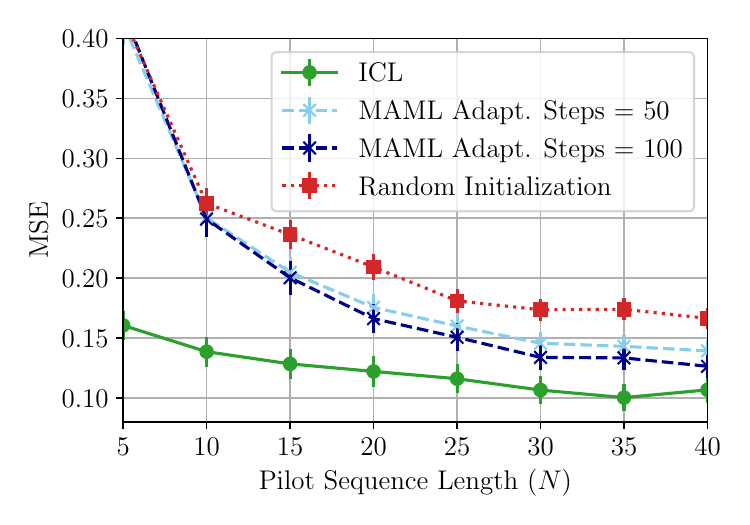}
	\vspace{-0.5cm}
 \caption{Test mean squared error of the ICL and MLP equalizers as a function of the pilot sequence length $N$.}
	\label{fig:maml_comp}
 	\vspace{-0.35cm}
\end{figure}

\noindent\emph{Comparison with MAML:}
In Figure \ref{fig:maml_comp}, we compare the performance of the ICL-based equalizer with that of a conventional multi-layer perceptron (MLP) equalizer. We specifically consider two MLPs, one that adapts to the pilot sequence starting from a random initialization and one that starts from a launch model obtained using MAML \cite{finn2017model}. ICL is observed to offer low MSE levels even at very short pilot lengths, while MAML requires a larger number of pilots to obtain a comparable performance. The downside of ICL is the scaling of the computational complexity with the number of pilots, denoted as $N$. In fact, due to the attention mechanism (\ref{eq:attention}), the complexity of ICL at test time increases with $N^2$, while that of MAML is generally linear in $N$, although it depends on the number of gradient steps during adaptation. For example, the complexity of MAML in our implementation was two times smaller than that of ICL in terms of floating-point operations when 50 gradient steps were implemented, while it matched that of ICL with 100 gradient steps.

\begin{figure}
	\centering
    \vspace{-0.1cm}
	\includegraphics[width=0.4\textwidth]{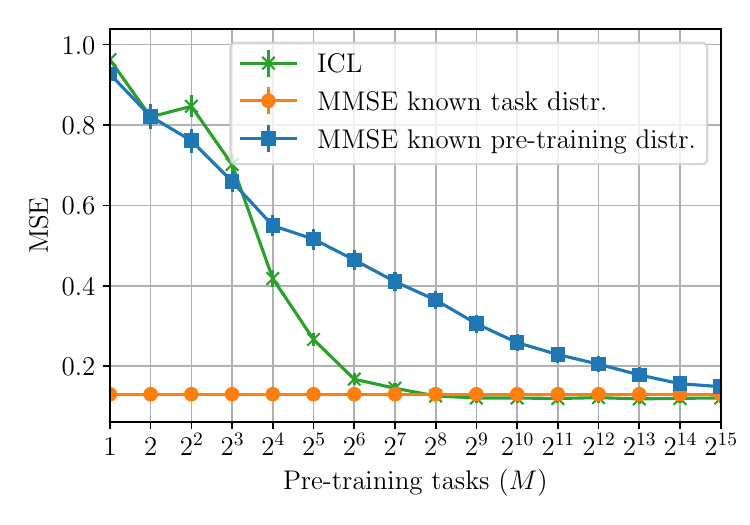}
	\vspace{-0.5cm}
 \caption{Test mean squared error of the ICL-based  and reference MMSE equalizers with different channel priors (Sec. \ref{subsec:knowndist}) as a function of the number of pre-training tasks $M$ ($b=4$).}
	\label{fig:msevstasks}
 	\vspace{-0.35cm}
\end{figure}

\noindent\emph{Threshold Behavior of ICL-Based Equalization:}
In this subsection, we study the generalization performance of the ICL-based equalizer as a function of the number of pre-training tasks, $M$, by considering a fixed noise power $\sigma^2=10$ dB, which corresponds to setting $\sigma^2_{\min}=\sigma^2_{\max}=0.1$. To this end, in Fig. \ref{fig:msevstasks}, we vary the size of the pre-training task set $M$ from $1$ to $2^{15}$, and we compare the performance of the ICL-based equalizer with the two MMSE equalizers with known channel distributions presented in Sec. \ref{subsec:knowndist}. 

The figure illustrates the anticipated threshold behavior akin to that revealed in \cite{raventos2023pretraining}. Specifically, for a small number of pre-training tasks $M$ -- here $M<2^5=32$ --, the ICL equalizer exhibits an MSE level similar to the MMSE estimator that uses as prior a uniform distribution over the  pre-training tasks' channel matrices. Accordingly, for a limited pre-training task diversity, the transformer cannot generalize to unseen tasks sampled from the channel distribution $P_H$. Conversely, for a sufficiently large number of pre-training tasks -- here $M>2^6=64$ --, the ICL equalizer is observed to extrapolate beyond the channel realizations seen during pre-training, aligning its performance with that of the MMSE estimator that uses as prior the true channel distribution $P_H$.

\begin{figure}
 \vspace{-0.1cm}
	\centering
	\includegraphics[width=0.4\textwidth]{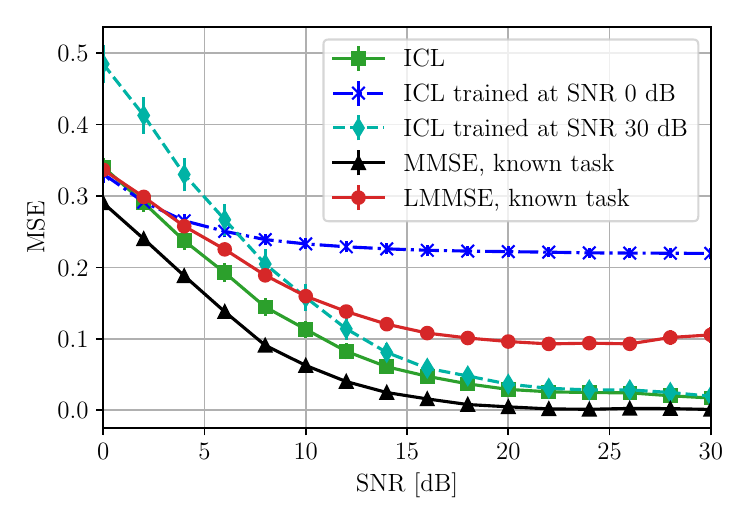}
 \vspace{-0.5cm}
	\caption{Test mean squared error as a function of the SNR level for the ICL-based equalizer trained at fixed SNR levels of 0 or 30  dB, as well as for the ICL-based equalizer trained on tasks with SNR levels uniformly drawn within the range $[0,30]$ dB. Also shown is the MSE of the benchmark  MMSE and LMMSE estimators with task knowledge presented in Sec. \ref{subsec:ideal} ($b=4$). } 
	\label{fig:msevssnr}
\end{figure}

\noindent\emph{Adaptivity to  SNR  Level:}
In the previous subsection, we focused on a task distribution characterized by a fixed noise level $\sigma^2$. In this subsection, we evaluate the capacity of the ICL-based equalizer to adapt to  diverse SNR levels. To this end, in Fig. \ref{fig:msevssnr} we plot the MSE as a function of the SNR for  ICL-based equalizers pre-trained using different task distributions. In particular, we consider ICL-based equalizers pre-trained at fixed SNR levels of 0 dB and 30 dB, as well as an ICL-based equalizer pre-trained using tasks with noise level uniformly sampled in the range $[-30,0]$ dB, corresponding to SNR levels $[0,30]$ dB. The figure includes also the reference MSE performance levels with known channel and SNR, as reviewed in Sec. \ref{subsec:ideal}.  

The ICL-based equalizers trained for a specific SNR level either outperform or match the performance of the LMMSE equalizer when tested on the same SNR level encountered during pre-training. This is despite the fact that, unlike the LMMSE equalizer, the ICL-based equalizers do not have access to the current channel realization. That said, training at a specific SNR level yields a significant MSE degradation when the SNR is different from the pre-training level. In stark contrast, the ICL-based equalizer pre-trained on tasks with diverse SNR levels uniformly outperforms the LMMSE benchmark, performing close to the ideal MMSE estimator. This illustrates the capacity of ICL to adapt to the true data distribution -- which does not follow the Gaussian assumption underlying LMMSE -- as well as to different SNR conditions.

\begin{figure}
	\centering
  \vspace{-0.3cm}
	\includegraphics[width=0.4\textwidth]{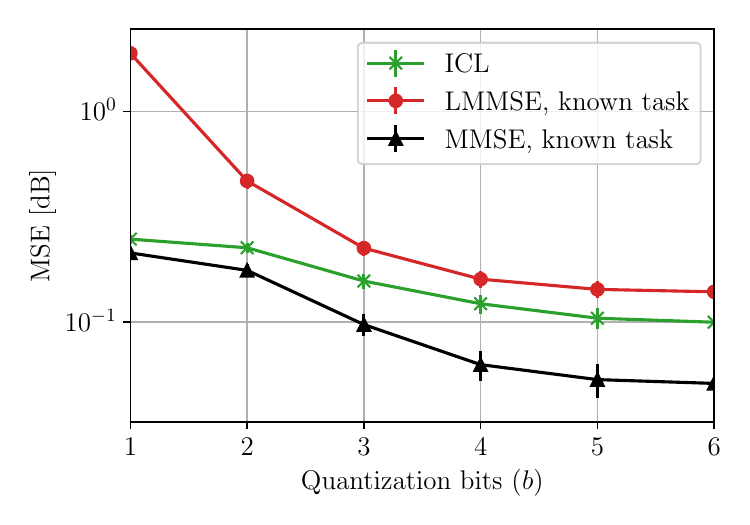}
 \vspace{-0.5cm}
	\caption{Test mean squared error as a function of the number of quantization bits $b$ for the ICL-based equalizer, as well as for the LMMSE and MMSE equalizers with task knowledge presented in Sec. \ref{subsec:ideal}.}
	\label{fig:msevsbits}
  \vspace{-0.35cm}
\end{figure}

\noindent\emph{Effect of Quantization:} In Fig. \ref{fig:msevsbits}, we evaluate the performance of the ICL-based equalizer, alongside the benchmarks MMSE and LMMSE estimators with task knowledge, as a function of the number of quantization bits $b$ used to quantize the received signal. Despite its knowledge of the channel matrix, the LMMSE estimator exhibits an MSE performance that degrades quickly as the number of bits $b$ decreases,   reflecting the limitations of linear equalizers in the presence of quantization.  In contrast, the ICL equalizer exhibits a more graceful performance degradation, mimicking the performance of the ideal MMSE equalizer. This indicates that the ICL-based equalizer can automatically learn to mitigate the non-linear distortion introduced by quantization.

\section{Conclusion}

In this paper, we have presented the idea of using a pre-trained transformer-based model to implement a direct mapping between an input determined by pilots and received data signals and an output given by equalized data symbols. Specifically, the equalizer produces a soft estimate of the transmitted signal via in-context learning (ICL) without requiring any explicit adaptation of its operation to changing channel conditions. One of the main conclusions of this study is that, given a sufficiently diverse data set during pre-training, ICL can approach the optimal MMSE equalizer, exhibiting a threshold behavior. Interesting open problems include the development of a generalization analysis and a study of ICL in the context of other applications of LLMs to communication systems  (see, e.g., \cite{wang2022transformer,maatouk2023large}).

\nocite{*}
\bibliographystyle{IEEEtran}
\bibliography{ref}

\end{document}